\documentclass[conference]{IEEEtran}

\usepackage[T1]{fontenc}
\usepackage[utf8]{inputenc}
\usepackage[english]{babel}
\usepackage{csquotes}
\usepackage{amsmath,amssymb}
\usepackage{booktabs}
\usepackage{graphicx}
\usepackage{float}
\usepackage{url}
\usepackage{multirow}
\usepackage{array}
\usepackage{xcolor}
\usepackage{microtype}
\usepackage{enumitem}
\usepackage{cite}
\usepackage{hyperref}

\graphicspath{{./}}

\title{Hagenberg Risk Management Process (Part 2): From Context-Sensitive Triage to Case Analysis With Bowtie and Bayesian Networks}

\author{
\IEEEauthorblockN{Eckehard Hermann \hspace{1.5em} Harald Lampesberger}
\IEEEauthorblockA{University of Applied Sciences Upper Austria, Campus Hagenberg, Austria\\
Email: \{eckehard.hermann, harald.lampesberger\}@fh-hagenberg.at}
}

\begin{document}

\maketitle

\begin{abstract}
Risk matrices (heatmaps) are widely used for information and cyber risk management and decision-making, yet they are often too coarse for today’s resilience-driven organizational and system landscapes. Likelihood and impact (the two dimensions represented in a heatmap) can vary with operational conditions, third-party dependencies, and the effectiveness of technical and organizational controls. At the same time, organizations cannot afford to analyze and operationalize every identified risk with equal depth using more sophisticated methods, telemetry, and real-time decision logic. We therefore propose a traceable triage pipeline that connects broad, context-sensitive screening with selective deep-dive analysis of material risks.

The Hagenberg Risk Management Process presented in this paper integrates three steps: (i) context-aware prioritization using multidimensional polar heatmaps to compare risks across multiple operational states, (ii) Bowtie analysis for triaged risks to structure causes, consequences, and barriers, and (iii) an automated transformation of Bowties into directed acyclic graphs as the structural basis for Bayesian networks. A distinctive feature is the explicit representation of barriers as \emph{activation nodes} in the resulting graph, making control points visible and preparing for later intervention and what-if analyses. The approach is demonstrated on an instant-payments gateway scenario in which a faulty production change under peak load leads to cascading degradation and transaction loss; DORA serves as the reference framework for resilience requirements. The result is an end-to-end, tool-supported workflow that improves transparency, auditability, and operational readiness from prioritization to monitoring-oriented models.
\end{abstract}

\begin{IEEEkeywords}
risk triage, operational resilience, context-aware risk assessment, Bowtie analysis, Bayesian networks, causal decision support
\end{IEEEkeywords}

\section{Introduction}

As today’s landscape grows more unpredictable, boosting \emph{resilience} has emerged as a key response~\cite{eu-resilience}.
The meaning of resilience varies somewhat depending on its subject or domain.
From a systems perspective, resilient systems are designed to anticipate disruptions, degrade, and recover without losing core services~\cite[p. 4]{hollnagel2006re}. In the resilience engineering literature, this is described as the ability to remain adaptive even under load (graceful extensibility)~\cite{woods2015four}. At an organizational level, resilience is understood as the ability to respond to change through leadership, culture, and governance mechanisms~\cite{iso22316}. In digital ecosystems and Information and Communication Technology (ICT), this definition extends even further, requiring risk and decision management to remain \emph{close to runtime}. Risks also change across contexts~\cite{rasmussen1997}, can unfold in cascades~\cite{buldyrev2010}, and are systematically influenced by technical as well as organizational controls~\cite{nist80039}.
Furthermore, regulatory requirements such as the Digital Operational Resilience Act (DORA)~\cite{dora2022} explicitly emphasize the need for resilient systems and processes, and it is reasonable to assume that other industries would likewise benefit from greater resilience.

Resilience directly affects how risks are managed, and the context dependency of risks~\cite{rasmussen1997} becomes a significant factor that we have also addressed in previous work~\cite{hermann2026polar}.
In the risk domain, heatmaps, i.\,e., risk matrices, are well established as tools for communication and prioritization and are regularly referenced in standards and guidelines (e.\,g., ISO~31000~\cite{iso31000} and ISO/IEC~31010~\cite{iec31010}).
However, for the systems mentioned above, a purely two-dimensional classification along likelihood and impact is often insufficient due to context dependency~\cite{rasmussen1997}: risks can vary across operating states, depend on third parties, and must be interpreted in relation to barriers and controls~\cite{nist80039}.
The notion of risk used here follows the scenario-based understanding of Kaplan and Garrick~\cite{kaplan1981}.
For the purposes of this paper, we consider three requirements for risk management to support resilience:
\begin{itemize}
  \item \textbf{Context dependency.} Risks must be assessed consistently across different operating states.
  \item \textbf{Control orientation.} The effectiveness of controls or barriers must be traceable.
  \item \textbf{Auditability.} Decisions, e.\,g., acceptance, treatment, and escalation, must be justified and documented reproducibly.
\end{itemize}

\subsection{Why triage is necessary}

At the same time, the practical reality remains that an organization must manage a very large number of risks at any given time.
Deep analysis, e.\,g., root-cause models, barriers, and evidence, typically relies on structured risk-analysis methods and barrier models~\cite{iec31010}. This requires expert time, data acquisition, ongoing model maintenance, alignment with operations and development, and often additional telemetry instrumentation~\cite{nist80030}. Additionally, monitoring incurs ongoing costs, e.\,g., signal acquisition, definition of service level objectives (SLOs), and alert handling, and requires established processes for continuous security monitoring~\cite{nist800137}. Embedding such measures into enterprise-wide risk management is typically addressed by established risk management frameworks~\cite{nist80039}. From a temporal and economic perspective it is therefore not possible to analyze \emph{all} risks in the same depth and transfer them into real-time capable monitoring and decision management. Instead, risk management standards emphasize prioritization and focused treatment of \emph{material risks}~\cite{iso31000}. Material risks are the ones that require continuous observation and short reaction times because their state can change quickly, early signals are eventually available, or cascades can lead to substantial impact~\cite{buldyrev2010}.

A practicable approach therefore requires an \emph{escalation and triage logic}: broad, rapid prioritization across contexts, followed by selective in-depth analysis only for those risks where real-time observation and short-term decision-making provide the greatest benefit.

\subsection{Contributions}

To address this challenge, this paper proposes our so-called \textit{Hagenberg Risk Management Process} consisting of three fundamental steps that we also address as pipeline:
\begin{enumerate}
  \item \textbf{Heatmap-based prioritization} with explicit contextual dimensions for broad initial assessment,
  \item \textbf{Bowtie modeling} of triaged material risks to structure causes, consequences, and barriers, and
  \item \textbf{DAG/Bayesian network (BN) modeling} for probabilistic operationalization and as a basis for intervention analysis, especially when monitoring and decision logic are coupled to event nodes and evidence~\cite{fenton2012}.
\end{enumerate}

Together, these steps form an integrated workflow from triage to actionable models that provide a basis for real-time monitoring, evidence updates, and what-if analyses for potential interventions. The paper addresses:

\begin{itemize}
  \item a recall of our context-sensitive risk modeling via multidimensional polar heatmaps for the triage of material risks \cite{hermann2026polar},
  \item a mapping of polar heatmap context factors onto Bowtie structures (e.\,g., events, forks, barriers),
  \item a ruleset for the projection of a Bowtie diagram into a directed acyclic graph (DAG) for representing a Bayesian network (BN) in line with recent work on pyBNBowTie~\cite{zurheide2021pybnbowtie} as well as related Bowtie$\rightarrow$BN approaches~\cite{khakzad2013},
  \item the marking of barriers as so-called \emph{activation nodes} and thus as potential intervention points with respect to causal modeling~\cite{pearl2009}.
\end{itemize}

This approach follows a design science methodology~\cite{hevner04design} grounded in the literature reviewed in Section~\ref{sec:related_work}. For evaluation, we introduce a use case in the following section and use it to illustrate the proposed steps. This paper concludes with a critical discussion of the evaluation, the study’s limitations, and an outlook on future work on intervention methods.

\section{Motivating Use Case: Instant Payments Gateway}
\label{sec:use_case}

Since DORA motivates this work, we define the following simplified scenario as a running example that is referred to in later sections for demonstrations.
Suppose a financial institution operates an instant payments gateway as a critical ICT service with an active/active architecture, i.\,e., two regions or availability zones serve traffic simultaneously.
A faulty production change, for example, a bad configuration parameter or routing rule, is rolled out during a peak-load window during business hours of the service.
Under load, the transaction queue builds up, clients experiencing delays begin to restart transactions (retry storm), and latencies increase, which ultimately result in partial to complete unavailability and eventually lost transactions.
Detection alone is insufficient because SLO-based alerts trigger when the crisis has already unfolded and signal quality is blurred by load artifacts (e.\,g., high but expected latency during business hours). Incident response starts with delay; a rollback is only performed after escalation (major incident), increasing overall impact.

The problem in this scenario can be summarized as: \textbf{The instant payments gateway outage is caused by a faulty change and insufficient detection under load.}

The scenario is DORA-relevant because it (i) affects a critical ICT service, (ii) addresses governance and control questions (change management, monitoring), and (iii) establishes a clear connection between controls and barriers and operational resilience~\cite{dora2022}.

\section{Related Work}
\label{sec:related_work}

\subsection{Heatmaps, risk matrices, and their limitations}
Risk matrices and heatmaps remain popular in organizational risk management because they compress multi-faceted judgments into a compact communication artifact. 
Risk matrices are intuitive, low-cost, and well suited as a communication artifact between business and engineering stakeholders. A risk is classified into discrete classes of likelihood and impact; color classes serve as acceptance and escalation rules~\cite{iso31000}. ISO/IEC~31010~\cite{iec31010} categorizes risk matrices as a risk analysis technique and describes typical forms of application. In practice, risk matrices are often used as \emph{risk acceptance criteria}, i.\,e., they define which risks are tolerated, mitigated, or escalated~\cite{duijm2015}. Especially in early phases (risk inventory, initial assessment), heatmaps are often the only instrument that both management and engineering stakeholders can apply consistently within an acceptable time.

However, Cox \cite{cox2008} identifies recurring shortcomings: coarse categorization can mask materially different risk profiles, ranking may become unstable, and seemingly precise color-coded outputs can invite overconfidence even when likelihood and impact scales are not commensurate. In this sense, matrices can be useful to structure discussion, but they are unreliable as a decision mechanism if their limitations are not made explicit. 

Duijm \cite{duijm2015} additionally discusses problems such as inconsistent scales, subjective class boundaries, and the risk that different risk magnitudes fall into the same color cell. This is particularly critical when risk matrices are misunderstood as hard acceptance limits although they only provide a coarse discretization. Moreover, according to Ball and Watt~\cite{ballwatt2013}, with unsuitable scaling a matrix can systematically misclassify risks, in particular when the color classes do not represent a monotonic mapping of a consistent risk concept.

In digital services, additional effects arise:
\begin{itemize}
  \item \textbf{Non-stationarity.} Likelihood and impact change dynamically (load, deployment, threat situation)~\cite{rasmussen1997}. A consistent organizational treatment of this dynamic requires an established enterprise risk management framework~\cite{nist80039}.
  \item \textbf{Dependencies.} Third-party services, networks, and platforms create coupled failure modes~\cite{buldyrev2010}.
  \item \textbf{Control effectiveness.} Barriers are not stably binary; their effectiveness depends on context (e.\,g., whether failover has been tested)~\cite{woods2015four}. As a resilience feature, we define this as the ability to remain adaptive even under load~\cite[pp. 4-5]{hollnagel2006re}.
\end{itemize}

This creates a tension: heatmaps are valuable as an entry artifact, but require (i)~a context-sensitive extension and (ii)~a defined handover to more advanced analysis techniques.
To address context-dependent systems, we previously introduced multidimensional polar heatmaps that represent multiple context factors and thresholds for state-dependent prioritization~\cite{hermann2026polar}. This paper builds on that screening capability and positions it as an explicit triage step, operationalizing the transition from context-sensitive prioritization (\emph{heatmap}) to causal and barrier-focused modelling (\emph{Bowtie}) and subsequent operational graph representations (\emph{DAG/BN}).

\subsection{Bowtie and barrier modeling}
Bowtie models integrate threat paths, a top event, and consequence paths in a single structure while making barriers/controls explicit. ISO/IEC~31010 \cite{iec31010} positions Bowtie as a technique in the ISO~31000 risk management toolbox and highlights the role of controls as the layer that links causal narratives to actionable measures. This makes Bowtie suitable as an intermediate artifact between organizational risk argumentation and operational risk control.

Moving beyond purely qualitative representations, de Dianous and Fi\'evez~\cite{dianous2006} (ARAMIS) treat barrier performance explicitly and interpret overall risk control as the combined effect of multiple barriers that may degrade or fail. This perspective is central for operational settings: controls are not only present or absent, but have a state that can change over time. In our approach, this motivates the explicit identification of barriers as observable and activatable elements that can later serve as operational intervention points.

\subsection{From Bowtie to Bayesian networks for operational updates}
For runtime assessments, evidence integration, and uncertainty propagation, a qualitative Bowtie model alone is insufficient. Khakzad et al.~\cite{khakzad2013} show how Bowtie structures can be mapped to Bayesian networks to enable probabilistic inference, incorporate evidence (e.\,g., from monitoring), and dynamically update the probabilities of the top event and consequences. A key contribution of this line of work is the probabilistic treatment of barriers (e.\,g. success/failure), which embeds barrier effectiveness directly into the risk assessment.

Zurheide et al.~\cite{zurheide2021pybnbowtie} complement this mapping with an implementation-oriented perspective by providing a Python library for Bayesian-network-based Bowtie analysis, addressing practical aspects such as transformation workflows, parametrization, and maintainability. Our contribution builds upon these foundations by introducing \emph{activation nodes} as explicitly marked intervention points in the derived DAG representation, designed to support deterministic model derivation and subsequent coupling to operational mechanisms.

\subsection{Causality and intervention}
While Bayesian networks support probabilistic inference and explanation, they do not, by themselves, justify claims about the effect of \emph{interventions}. Koller and Friedman~\cite{koller2009pgm} provide the methodological foundations for probabilistic graphical models, including structuring and inference, which underpin the BN/DAG layer in this work.

Pearl \cite{pearl2009} formalizes causal reasoning and the do-calculus, clarifying the difference between observational statements and interventional questions of the form \enquote{What happens if we do X?}. This distinction is crucial for resilient operational settings: monitoring produces evidence, but control requires deliberate actions. In our approach, barriers are therefore modeled as explicit intervention points, providing a principled rationale for treating \emph{activation nodes} as the bridge between risk detection and operational decision management.

\section{Multidimensional polar heatmaps}
\label{sec:heatmaps}

\subsection{Definition and classification of context factors}
In our proposed polar heatmap approach for risk triage, context factors are conditions that systematically change the likelihood or impact of a risk without themselves being the top event. We distinguish:
\begin{itemize}
  \item \textbf{X context (likelihood):} factors that change the likelihood of occurrence (e.\,g., change complexity, third-party health).
  \item \textbf{Y context (impact):} factors that change the damage after occurrence (e.\,g., failover maturity, traffic criticality).
\end{itemize}
In the discussed use case, \emph{Operational Load Context} is particularly relevant as an additional dimension, as transactional load affects both likelihood (instability) and impact (backlog, timeouts, lost transactions).

\subsection{Formal model: ND heatmap and ND slices}
We model a risk object $r$ by a base heatmap $H_r$ over two main axes:
\[
H_r: \mathcal{X} \times \mathcal{Y} \rightarrow \mathcal{R},
\]
where $\mathcal{X}$ (likelihood level) and $\mathcal{Y}$ (impact level) are scales and $\mathcal{R}$ denotes a risk classification, e.\,g., color value, score, or acceptance class. Context dimensions $C_1,\dots,C_n$ extend the model to a tensor:
\[
T_r: \mathcal{X} \times \mathcal{Y} \times \mathcal{C}_1 \times \cdots \times \mathcal{C}_n \rightarrow \mathcal{R}.
\]
An $n$-dimensional slice (\emph{ND slice}) is obtained by indexing (slicing) all context dimensions at a specific state $\mathbf{c}=(c_1,\dots,c_n)$:
\[
S_{r,\mathbf{c}}(x,y) = T_r(x,y,c_1,\dots,c_n).
\]
ND modeling is particularly useful when risks differ structurally across operating situations: in that case, not only $x,y$ vary, but the acceptable risk threshold may also depend on context.

\subsection{Polar heatmap as visualization}

An ND slice can furthermore be visualized as a polar heatmap~\cite{hermann2026polar}: Each context dimension is assigned to an angular range, each level to a radial distance. In practice, this produces a \emph{contextual signature} that shows the selected levels at a glance, while the color coding reflects the resulting risk grade. The polar heatmap thus extends the classical 2D heatmap with the explanatory component \enquote{which context leads to the position in the risk space}. A practical advantage is reduced cognitive load: stakeholders can (i) understand the contextual configuration and (ii) see the classification (color) in the same artifact.

\subsection{Context-based adjustments}
In many organizations, risk estimates arise as baseline estimates that are further adjusted by context factors. For the tool operationalization in the considered use case in this paper, we use a simple, auditable addition scheme:
\begin{align}
X_{\text{adj}} &= X_{\text{base}} + \sum_{i} \Delta X_i,\\
Y_{\text{adj}} &= \max\left(0,\; Y_{\text{base}} + \sum_{j} \Delta Y_j\right),
\end{align}
where $\Delta X_i$ and $\Delta Y_j$ are context-dependent adjustment values for likelihood and impact, respectively, e.\,g., peak load in our use case increases both likelihood $\Delta X>0$ and impact $\Delta Y>0$ of a bad outcome. This scheme primarily enables prioritization and is complemented in the Bowtie/DAG step by structured causal models. It is kept intentionally simple because it can be communicated transparently and is well suited for governance (e.\,g., change advisory, risk acceptance).

\begin{figure*}[t]
  \centering
  \includegraphics[width=0.98\textwidth]{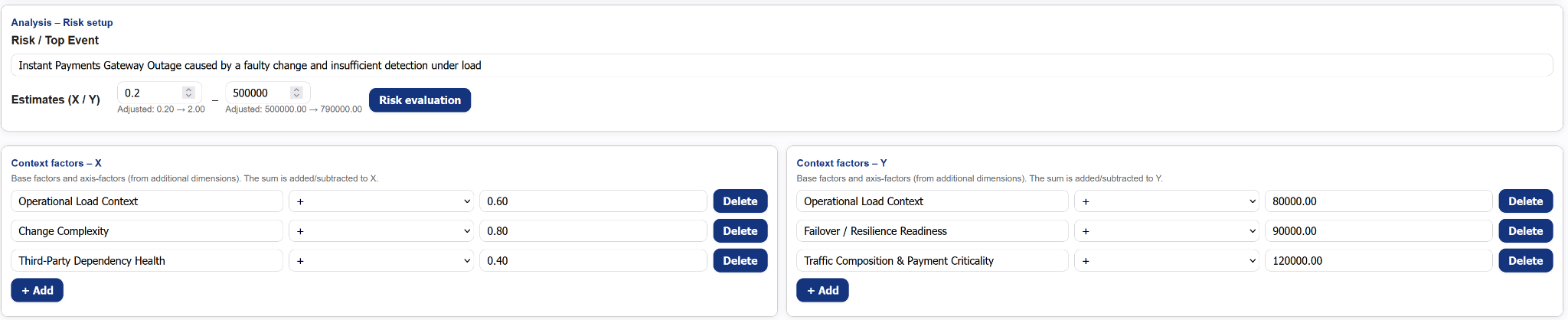}
  \caption{Risk editor of the Risk Analyser: assignment of several context factors that influence likelihood (X axis) and transactions (Y axis).}
  \label{fig:heatmap2d}
\end{figure*}

\begin{figure*}[t]
  \centering
  \includegraphics[width=0.98\textwidth]{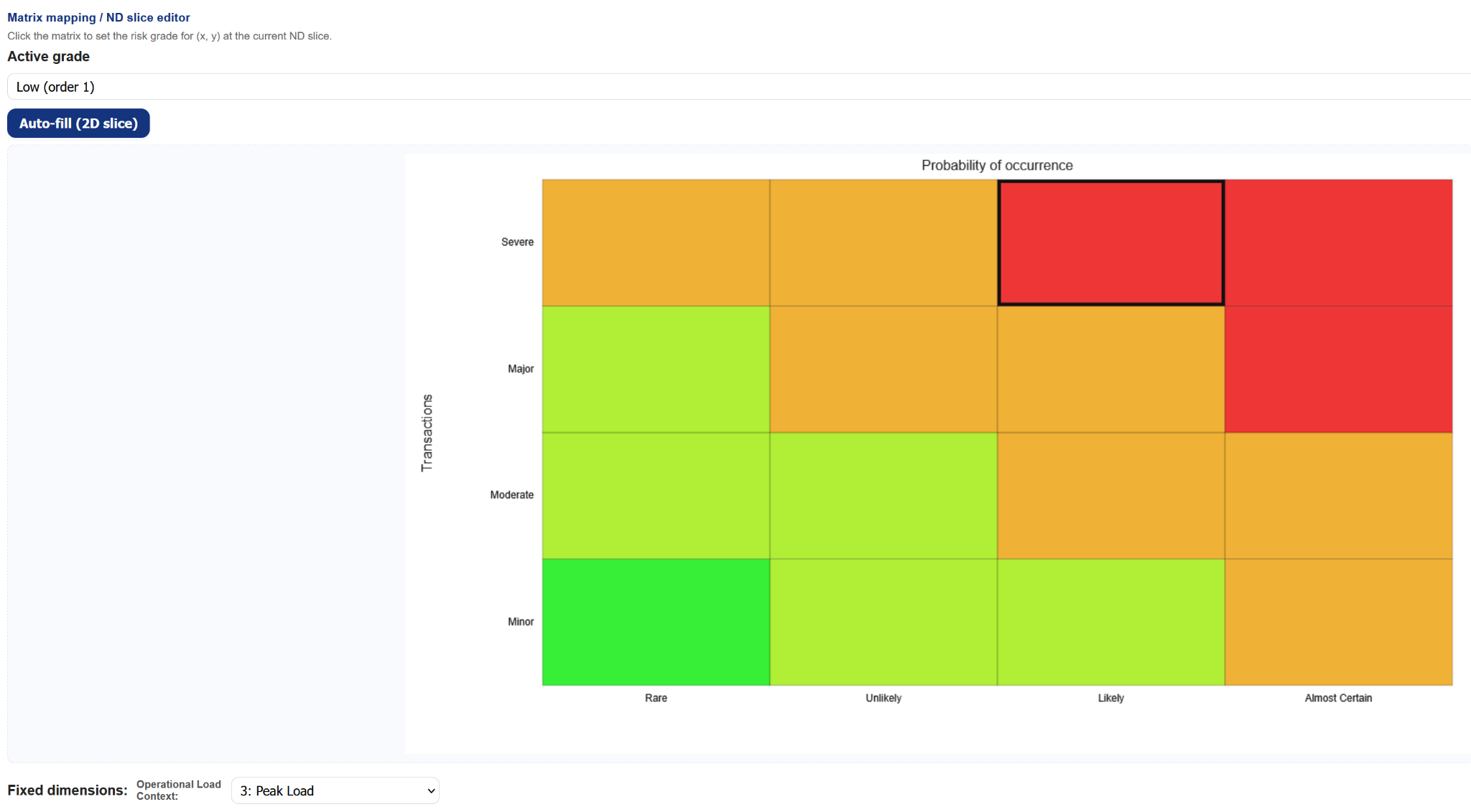}
  \caption{ND slice of a heatmap: Fixed dimensions show the selected level of the Operational Load Context axis.}
  \label{fig:polar}
\end{figure*}

\begin{figure}[t]
  \centering
  \includegraphics[width=\columnwidth]{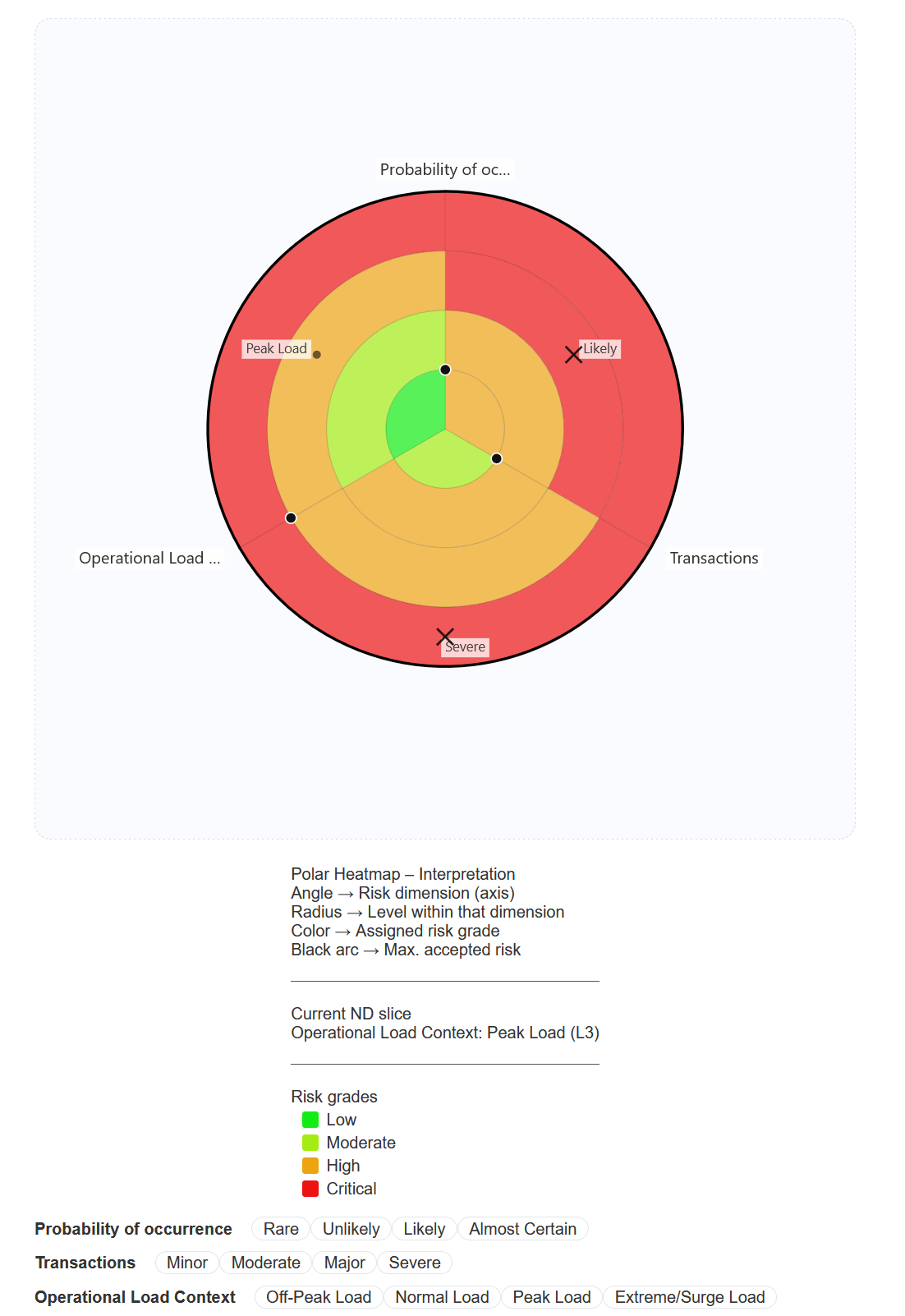}
  \caption{Polar heatmap with legend/interpretation.}
  \label{fig:polarlegend}
\end{figure}

To demonstrate visualizations based on the discussed use case in this paper, Fig.~\ref{fig:heatmap2d} shows the assignment in the implemented slice editor for a single, fixed context (Operational Load Context = Peak Load). Fig.~\ref{fig:polar} shows the corresponding 2D heatmap slice and Fig.~\ref{fig:polarlegend} the corresponding polar heatmap.

\subsection{Design of axes, levels, and thresholds}
The quality of a heatmap depends substantially on the choice of axes, their scaling, and the definition of level boundaries. The literature points out that inconsistent or non-monotonic scales are key weaknesses of risk matrices~\cite{cox2008}. Duijm additionally discusses design recommendations for the design and interpretation of risk matrices~\cite{duijm2015}. For a practical and auditable application, we distill the insights from the literature into three guidelines:

\begin{enumerate}
    \item \textbf{Define semantics per axis unambiguously.} Each axis must map a clearly interpretable quantity (e.\,g., \enquote{lost transactions} instead of \enquote{high impact}).\\
    \item \textbf{Couple boundaries to decision or operating rules.} Level boundaries should, as far as possible, be coupled to thresholds from SLOs, capacity planning, or regulatory reporting/escalation rules.\\
    \item \textbf{Choose levels so that an ordering is stable.} Especially for ordinal likelihood scales, the classes should have a robust meaning that is understandable in everyday practice, e.\,g., frequency per time interval.
\end{enumerate}

\begin{table}
\centering
\caption{Dimensions and level definitions for our use case.}
\label{tab:xybounds}
\begin{tabular}{lll}
\toprule
\textbf{Dimension} &  \textbf{Scale} & \textbf{Levels (name and range)} \\
\midrule
$X$ Likelihood & ordinal & \textit{Rare}: under 1/year;\\
 &  & \textit{Unlikely}: $1$--$2$/year;\\
&  & \textit{Likely}: $3$--$12$/year;\\
 &  & \textit{Almost certain}: over $12$/year.\\[0.5em]
 
$Y$ Impact (lost transactions) & ordinal & \textit{Minor}: $0$--$10$;\\
&  & \textit{Moderate}: $11$--$100$;\\
 &  & \textit{Major}: $101$--$1{,}000$;\\
&  & \textit{Severe}: over $1{,}000$. \\[0.5em]
Operational Load Context & ordinal & \textit{Off-Peak}: under $40\%$;\\
 &  & \textit{Normal}: $40$--$70\%$;\\
 &  & \textit{Peak}: $70$--$90\%$;\\
 &  & \textit{Surge}: over $90\%$ / overload. \\
\bottomrule
\end{tabular}
\end{table}

\subsection{Example: axes and boundaries in the use case}
To model the use case, we use for $Y$ a metric impact axis reflecting \emph{lost transactions} and for $X$ an ordinalized likelihood axis that is oriented towards expected frequency in an observation period. Table~\ref{tab:xybounds} shows the dimensions and gives example boundaries and level names which, in practice, must be adapted to the institution's transaction volumes.

For the additional context axis \emph{Operational Load Context}, relative capacity limits are particularly suitable because they remain comparable across different absolute scales. A simple operationalization uses the ratio of current load to sustainably available capacity (incl.\ reserves) as shown in Table~\ref{tab:xybounds}.
These boundaries can be connected directly to capacity planning, autoscaling rules, and SLO budgets and are thus auditable.

\subsection{Relationship between context axes and X/Y}
Context axes are not a third dimension without influence; rather, they serve to model systematic shifts of $X$ and $Y$. In the heatmap phase, this is done via modeling adjustments as considered in Equations (1)--(2). In the Bowtie/DAG step, the relationship is resolved structurally: context factors become cause or impact variables, and barriers capture \emph{how} and \emph{where} the shift arises or can be compensated. This yields a consistent chain from semi-quantitative prioritization to an explanatory model structure.

\section{From polar heatmap to Bowtie analysis}
Heatmaps serve coarse prioritization. For prioritized risks, a model of the causal and consequence side is required that makes controls explicit. ISO/IEC~31010~\cite{iec31010} introduces Bowtie diagrams as a technique for risk analysis, in particular for representing causes, top event, consequences, and barriers. Bowties are therefore suitable as the next analysis step for risks that fall into non-acceptable areas in the polar heatmap.

\subsection{Triage: from broad scan to real-time capability}
The polar heatmap supports \emph{broad} prioritization across many contexts. The transition to Bowtie analysis deliberately marks a \emph{triage decision}: only a subset of risks is further studied in depth, because full operationalization, i.\,e., detailed analysis, data integration, monitoring, and decision logic, does not scale. Criteria for \emph{material risks} are in particular (i) exceeding acceptance limits in at least one relevant context and (ii) high \emph{temporal criticality} (i.\,e., short time windows between first signs and high impact). For such risks, in-depth analysis is worthwhile because observability (evidence), barriers, and intervention options can be anchored in a model and transferred into runtime processes. Risks with low temporal dynamics or long reaction windows, by contrast, remain in periodic reviews and action portfolios, without permanent monitoring being economically justified.

\subsection{Why Bowtie after the heatmap?}
The polar heatmap primarily answers: How critical is the risk in a context? A Bowtie model complements this with the questions:
\begin{itemize}
  \item {Why} does the risk occur (threats/causes)?
  \item {Which} consequence paths are possible, or how can consequences evolve?
  \item {Which} controls (barriers) exist and where do they act?
\end{itemize}
This structure is crucial not only for listing measures, but for anchoring them at specific cause or impact points. This is particularly important in regulated settings such as DORA, where controls and resilience measures must be justifiable within governance and audits, for example, which evidence for a given control class was considered and which risks are addressed.

\subsection{Mapping of context factors}
\begin{figure*}[t]
  \centering
  \includegraphics[width=0.99\textwidth]{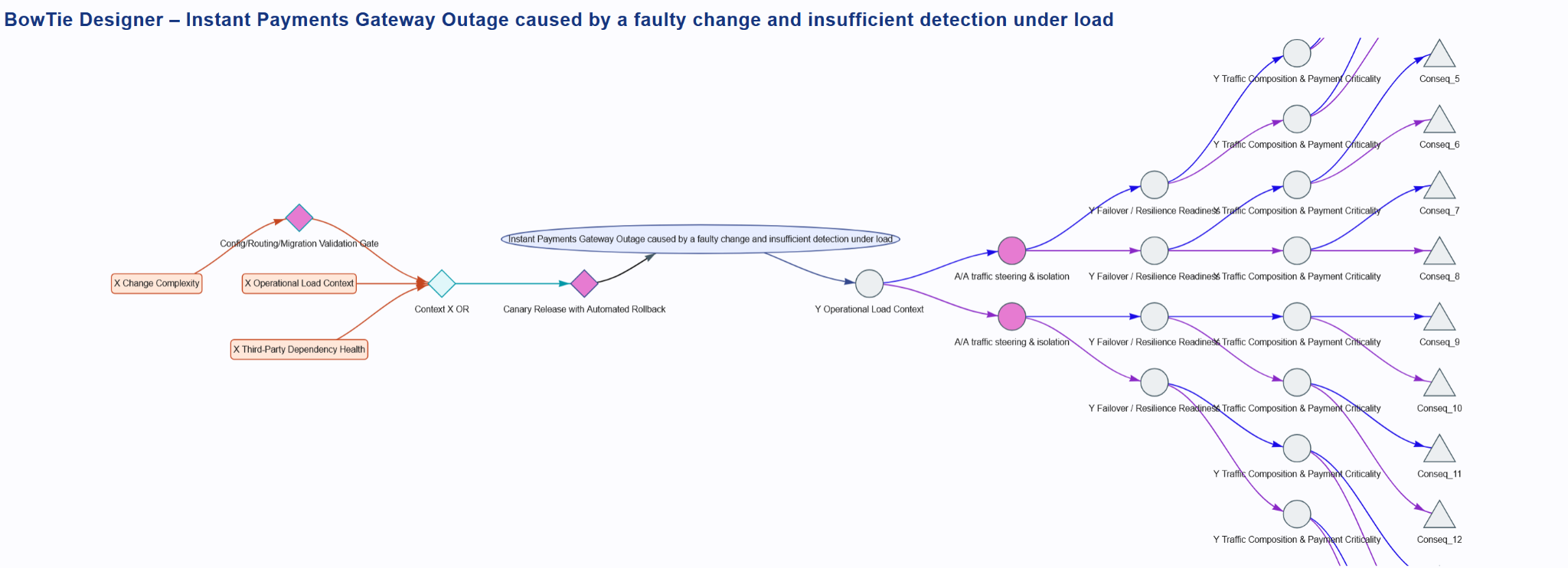}
  \caption{Bowtie with context factors mapped onto barriers (fault tree left, event tree right).}
  \label{fig:bowtie}
\end{figure*}

Our implemented approach integrates ND heatmap context factors as structuring elements in the Bowtie:
\begin{itemize}
  \item \textbf{X context (likelihood):} context factors that influence likelihood are inserted on the left as cause nodes and integrated into the top-event logic via an OR gate.
  \item \textbf{Y context (impact):} context factors that shape the impact after occurrence are inserted on the right as a directed chain starting from the top event and influence the consequence paths.
\end{itemize}
This modeling has two advantages: context is explicitly integrated into the cause/impact structure and the Bowtie$\rightarrow$DAG mapping projects this structure without semantic loss.
Fig.~\ref{fig:bowtie} shows the generated Bowtie for the use case and context factors denoted in Fig.~\ref{fig:heatmap2d}.

\subsection{Bowtie structure and barrier semantics}
A Bowtie connects cause paths on the left-hand side (fault-tree-like) with a central top event and models consequence paths on the right-hand side (event-tree-like). In the Bowtie literature, the benefit is seen in particular in making barriers explicit: de Dianous and Fiévez~\cite{dianous2006} show in the ARAMIS context how Bowties are used to demonstrate risk control and to assess barrier performance.

In an ICT context, this perspective is analogous: barriers correspond to controls such as change gates, canary rollbacks, isolation, or traffic steering. They can be classified into classes:
\begin{itemize}
  \item \textbf{preventive:} to prevent occurrence; e.\,g., change gates, canary rollback,
  \item \textbf{detective:} to shorten the Mean Time To Detect (MTTD); e.\,g., SLO alerts, synthetic checks,
  \item \textbf{mitigative:} to reduce impact; e.\,g., traffic steering, circuit breaker, and
  \item \textbf{corrective:} to shorten the Mean Time to Recovery (MTTR); e.\,g., runbooks, automation.
\end{itemize}
In the use case, the focus is on preventive and mitigative barriers because the scenario affects both occurrence and damage.

\subsection{Positioning of barriers}
With respect to the use case, in active/active architectures, positioning of barriers is central in order to avoid redundant barrier stacks and instead cover different failure modes:
\begin{itemize}
  \item {FT-1 (preventive, change-specific):} \emph{Config/Routing/Migration Validation Gate} between change complexity and the X-context OR gate.
  \item {FT-2 (preventive, system-wide):} \emph{Canary Release with Automated Rollback} between the X-context OR gate and the top event.
  \item {ET-1 (mitigative, Active/Active):} \emph{Health-based traffic steering \& isolation} directly after the top event.
\end{itemize}
Thus, FT-1 addresses error creation (shift-left), FT-2 addresses early correction under real load, and ET-1 addresses impact reduction through distributed routing and cascade avoidance.


In classical Bowtie diagrams, barriers are often documented as \enquote{present/not present} or assessed qualitatively. For further processing in probabilistic models, it is advantageous to model barriers as their own state variables with success/failure mode~\cite{khakzad2013}. This allows sensitivity analyses (e.\,g., ranking by risk reduction) and makes dependencies visible when barriers are not independent (e.\,g., shared observability stack).

\subsection{Activation nodes}
We designate barriers as \emph{activation nodes}, i.\,e., nodes that are semantically considered as intervenable (e.\,g., works/fails). Formally, an activation node $A_k$ is a variable whose state can be actively set in analyses in order to model measures. In the present contribution, this property is first modeled as \emph{metadata} on the node so that later analyses can unambiguously identify the set of intervenable nodes. In a follow-up contribution (Hagenberg Risk Management Process (Part 3)), this semantic is elaborated as an intervention operator in causal graphical models~\cite{pearl2009}.

\section{Bowtie$\rightarrow$DAG Transformation}

\subsection{Bayesian networks}
A directed acyclic graph (DAG) is a directed graph of nodes and edges without cycles. DAGs are the structural basis of Bayesian networks and their factorization:
\[
P(V_1, V_2, \dots, V_n) = \prod_{i = 1}^{n} P\left(V_i \mid \mathrm{Pa}(V_i)\right),
\]
where $\mathrm{Pa}(V_i)$ are the parent nodes of $V_i$~\cite[p. 62]{koller2009pgm}.
This factorization allows uncertainty and dependencies to be operationally modeled and inference to be performed.

\subsection{Mapping rules: Bowtie $\rightarrow$ DAG}
pyBNBowTie \cite{zurheide2021pybnbowtie} describes a Bowtie$\rightarrow$BN mapping and provides an implementation. In the process-safety literature, related mapping approaches exist that transfer Bowties into Bayesian networks in order to model dependencies and updates~\cite{khakzad2013}. The tool used here follows the same basic principles and extends them to ICT Bowties with context factors.

Neither (polar) heatmaps nor Bowtie diagrams are suitable as a model basis for operational monitoring, because they do not explicitly represent a system's normal state: heatmaps provide a context-related prioritization of risks, and Bowties structure causes, top event, and consequences of the risk event, but they always start from the event. In monitoring, by contrast, normal operation dominates, in which the top event has not occurred and the system operates \enquote{safe}. A DAG, or a Bayesian network based on it, is therefore required for a probabilistic representation of this normal state and for inference from observable evidence. Accordingly, the Bowtie$\rightarrow$DAG mapping in pyBNBowTie supplements the consequence side with an additional \enquote{safe} state that models the normal state and thus makes probabilities for \enquote{safe} vs.\ adverse consequences accessible in the Bayesian network~\cite{zurheide2021pybnbowtie}.
The Bowtie $\rightarrow$ DAG transformation rules can be summarized as follows:
\begin{itemize}
  \item Event and consequence nodes become BN variables (binary or multi-state).
  \item Gates (AND/OR) are modeled as deterministic nodes. Example OR: $G=\bigvee_i P_i$.
  \item Barriers are inserted as their own nodes that, as parents, modulate the downstream event node (barrier works reduces occurrence).
  \item Event-tree branches are represented by fork nodes that structure alternative consequence paths.
\end{itemize}
The determinism can be extended as required by leak probabilities and uncertainty models; in the present contribution, the focus is on structural transfer.

\subsection{From deterministic to probabilistic nodes}
For quantitative evaluation, it is common to transfer deterministic gate semantics into BN probabilities, where uncertainty or incomplete information needs to be modeled explicitly. A widespread simplification is the \emph{Noisy-OR} mechanism for OR-like relationships. For a node $E$ with parents $P_1,\dots,P_k$ and independent influence parameters $\theta_i$, this yields:
\[
P(E=1 \mid P_1,\dots,P_k)=1-\prod_{i: P_i=1}(1-\theta_i).
\]
Barriers can be modeled as additional parents that dampen the node activation. For a binary barrier $B \in \{\text{works},\text{fails}\}$, one can, for example, use a damping factor $\lambda \in [0,1]$:
\begin{multline*}
P(E=1 \mid P_1,\dots,P_k, B=\text{works})=\\
\lambda \cdot P(E=1 \mid P_1,\dots,P_k, B=\text{fails}) .
\end{multline*}
Such formulations are compatible with Bowtie$\rightarrow$BN approaches \cite{khakzad2013} and can be stored as parameter templates in the tool implementation. Importantly, the \emph{structure} (which nodes influence which events) is already fixed by the DAG; parameterization thus becomes a separate, transparent activity.

\subsection{Activation nodes and what-if analyses}
Activation nodes mark precisely those variables that are \enquote{switchable} from the perspective of measures. Even without exact parameterization, qualitative what-if questions can be answered by setting node values and structurally tracing consequence paths:
\begin{itemize}[leftmargin=*,nosep]
  \item If \emph{Canary rollback works}, the probability that a faulty change escalates to the top event decreases.
  \item If \emph{Health-based steering works}, the consequence path towards \emph{lost transactions} is shortened or weakened (e.\,g., by isolating a region).
\end{itemize}
In the follow-up contribution Hagenberg Risk Management Process (Part 3), these interventions are formalized via the do-operator and combined with Bayesian inference~\cite{pearl2009}.

\subsection{Monitoring and decision management}
\begin{figure*}[t]
  \centering
  \includegraphics[width=0.98\textwidth]{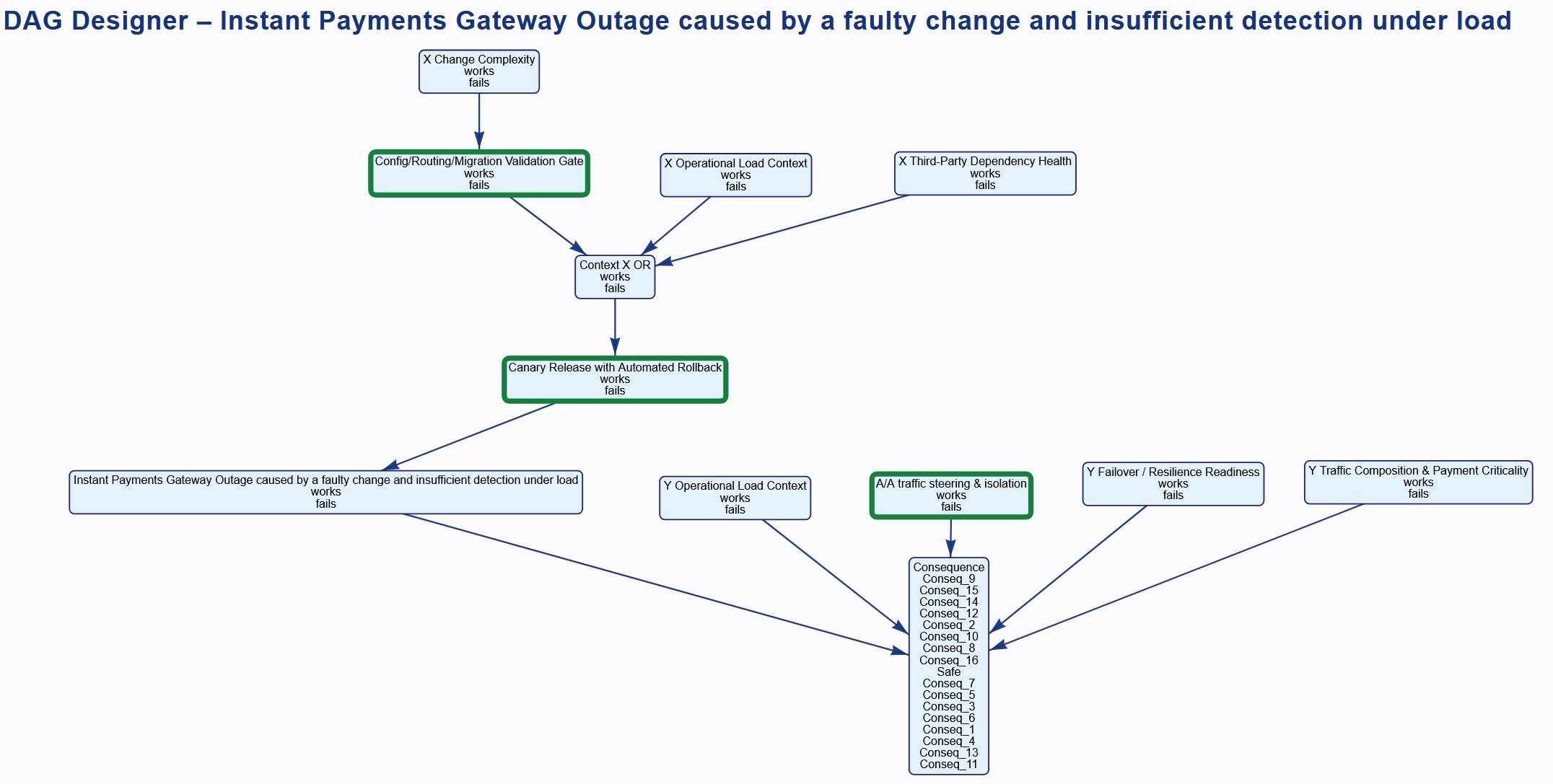}
  \caption{DAG derived from Bowtie with marked barriers as activation nodes.}
  \label{fig:dag}
\end{figure*}

A DAG expresses assumptions about dependencies explicitly and creates the prerequisite for quantitative statements (posteriors, evidence, sensitivity). At the same time, a DAG can serve as a common structure for monitoring and decision support by interpreting telemetry as observation of nodes. For DORA-near environments, this is relevant because evidence of control effectiveness is often based on measurement data (observability).
For visualizing the Bowtie$\to$DAG transformation, Fig.~\ref{fig:dag} shows the DAG derived from the Bowtie in Fig.~\ref{fig:bowtie}, including the marked activation nodes.

\section{Evaluation: Instant Payments Gateway}

To evaluate the proposed approach, the pipeline was implemented in a tool that manages polar heatmaps, Bowties, and DAGs in a consistent model. For the scientific argument, two properties are central: (i) applicability of the proposed steps, (ii) persistence of the model artifacts (axes, context dimensions, Bowtie structure) and (iii) the deterministic mapping layer Bowtie$\rightarrow$DAG, which yields reproducible results.

\subsection{Data model and persistence}
Risk objects contain a base heatmap (X/Y), optionally additional context axes, and rules for ND slices. The polar heatmap is generated from the ND slice and shows, per context dimension, the level expression. For Bowties, nodes (threats, gates, barriers, consequences) and edges are persistently stored; barriers additionally carry a binary state model (works/fails) and are marked as activation nodes.

\subsection{Use-case-specific context adjustments}
For heatmap prioritization in the use case, context-dependent adjustments were implemented as simple additive contributions. Table~\ref{tab:deltas} shows a consistent, auditable set of example contributions for a total of four context factors that are also visible in Fig.~\ref{fig:heatmap2d}, Fig.~\ref{fig:bowtie}, and Fig.~\ref{fig:dag}.
$\Delta Y$ is estimated from an assumed additional loss rate $r_{\text{loss}}$ and exposure time $t$ over an incident window $T=10$~min, consistent with the $Y$ bins in Table~\ref{tab:xybounds}.

\begin{table*}
\centering
\small
\setlength{\tabcolsep}{4.5pt}
\caption{Example context contributions $\Delta X$ and $\Delta Y$ in the use case.}
\label{tab:deltas}
\begin{tabular}{@{}lrrrr@{}}
\toprule
\textbf{Context factor (level)} & $\Delta X$ & $r_{\text{loss}}$ [txns/min] & $t$ [min] & $\Delta Y$ [txns] \\
\midrule
Operational load (peak) & +0.6 & 120 & 3 & 360 \\
Change complexity (high) & +0.8 & 90 & 2 & 180 \\
Third-party dependency health (degraded) & +0.4 & 110 & 3 & 330 \\
Other (baseline disturbance) & +0.0 & 15 & 2 & 30 \\
\midrule
\textbf{Total} & \textbf{+1.8} &  & \textbf{10} & \textbf{$Y=900$} \\
\bottomrule
\end{tabular}
\end{table*}

\subsection{Heatmap and polar-heatmap phase: prioritization}
The heatmap phase provides rapid, but documentable prioritization. Fig.~\ref{fig:heatmap2d} shows the slice editor: for a fixed context state (Operational Load Context = Peak Load), a point $(X,Y)$ is determined in the 2D plane, where $Y$ is modeled as \emph{lost transactions}. The level boundaries for $X$ and $Y$ follow Table~\ref{tab:xybounds}; they are chosen so that they (i) are intuitively interpretable for stakeholders and (ii) can be coupled to operating rules (capacity planning, SLOs, escalation).

The polar heatmap (Fig.~\ref{fig:polar}--\ref{fig:polarlegend}) represents the context state transparently. In the slice shown, load is configured as peak; this increases both likelihood (e.\,g., increased instability under configuration errors) and impact (high throughput, thus high potential loss counts). In the tool operationalization, this mechanism is represented by context-dependent adjustments (cf.\ Table~\ref{tab:deltas}): operational load, change complexity, and third-party health can, if applicable, affect $X$ and $Y$ in different directions and with different magnitudes.

\textbf{Example slice calculation.} Assume the baseline estimate is $X_{\text{base}}=\text{Unlikely}$ and $Y_{\text{base}}=\text{Moderate}$. For a peak-load state (+2/+2), $X_{\text{adj}}$ shifts towards \emph{Likely} to \emph{Almost certain} and $Y_{\text{adj}}$ towards \emph{Major} to \emph{Severe}. This yields a clear, justified escalation into the non-acceptable area without the rationale being \enquote{hidden} in free text.

However, this phase deliberately does not answer causal questions in detail; it is a prioritizing filter. Nonetheless, the key benefit is the reproducible documentation of the context state that leads to escalation. Exactly this context state is modeled into the Bowtie structure in the next step.

\subsection{Bowtie phase: structuring and barriers}
The Bowtie (Fig.~\ref{fig:bowtie}) makes cause paths (fault tree) and impact evolution (event tree) concrete for the prioritized context state. It becomes visible that the likelihood side is substantially determined by change-related factors and dependency states, while the impact chain is shaped by load state and resilience measures (traffic steering, isolation, circuit breaker). The explicit positioning of barriers also shows a clear control logic: prevention reduces the probability of the top event, mitigation reduces impact after occurrence.

\subsection{DAG phase: operationalization}

The Bowtie$\rightarrow$DAG mapping follows a deterministic transformation sequence building upon pyBNBowTie \cite{zurheide2021pybnbowtie}:
\begin{enumerate}
  \item Normalization of the Bowtie structure (gates explicit, barriers as their own nodes).
  \item Transfer of all nodes into DAG nodes; preservation of edge direction (cause $\rightarrow$ effect).
  \item Replacement of logical gates by deterministic CPTs (OR/AND).
  \item Insertion of barriers as parent nodes of downstream event nodes; marking as activation nodes.
  \item Linearization/explicit representation of event-tree sequences using fork/sequence nodes so that alternative consequence paths are represented acyclically and unambiguously in the DAG.
\end{enumerate}
These steps preserve structure: every Bowtie interpretation (causes, top event, consequences, barriers) remains recognizable.
The DAG (Fig.~\ref{fig:dag}) makes dependencies and barrier states explicit and prepares probabilistic analyses (e.\,g., evidence from monitoring). The marking of barriers as activation nodes is the central transition: it creates an unambiguous set of nodes that can be actively set as measure states (e.\,g., \enquote{Canary rollback works}) in order to investigate the effect of controls on top event and consequences quantitatively.

\subsection{DORA Context}
DORA \cite{dora2022} addresses digital operational resilience along the entire lifecycle of critical ICT services: identification and management of ICT risks, incident handling and reporting, resilience testing, and management of third parties. For the model workflow considered in this paper, the following technical and methodological requirements arise in particular.

DORA demands explicit consideration of criticality and dependencies. For instant-payments services, context states such as peak load or change windows are essential: an identical technical problem may be tolerable off-peak, but non-acceptable at peak. Multidimensional polar heatmaps \cite{hermann2026polar} support this requirement by modeling context as an axis and carrying it along with each assessment.

DORA requires not only the existence of controls, but traceable embedding into processes and evidence of effectiveness. The Bowtie analysis provides a structured view for this: preventive controls are positioned in the fault tree, mitigative controls in the event tree. By transferring into a DAG, barriers can be modeled as their own variables so that effectiveness can be examined not only qualitatively, but (in follow-up work) quantitatively.

In the use case, \enquote{insufficient detection under load} is central. This underlines that observable signals, such as SLO alerts or health signals, are not only technical telemetry, but can be interpreted as evidence with respect to a probabilistic model. The DAG step enables such evidence to be structurally attached to cause or barrier nodes, supporting reproducibility of analyses in incident review.

\section{Threats to validity and limitations}
The proposed pipeline is to be understood as a methodological framework and tool implementation. The following restrictions are essential for interpreting the results:
\begin{itemize}
  \item {Semi-quantitative prioritization:} The heatmap phase works with discrete levels and simple adjustments. This is deliberately chosen, but does not replace a quantitative analysis in case of material risks.
  \item {Parameterization of the BN:} The transformation procedure from polar heatmap to Bowtie to DAG provides a grounded structure, but no robust probabilities. These must be derived and validated from data, expert knowledge, or tests.
  \item {Dependency assumptions:} Gate determinism and independence assumptions are modeling decisions in the presented procedures. They are transparent, but must be checked in the respective context.
\end{itemize}
These limits motivate the next step: a more formal treatment of activation nodes as intervention points as well as data-driven parameterization.

\section{Discussion}
From a governance and audit perspective, the proposed pipeline of the Hagenberg Risk Management Process provides clear traceability:
\emph{context state} $\rightarrow$ \emph{heatmap position} $\rightarrow$ \emph{Bowtie structure} $\rightarrow$ \emph{DAG structure}. This workflow makes it transparent why a risk is escalated under certain conditions, which barriers are relevant, and how they are structurally transferred into a quantitative analysis. This traceability is a decisive advantage compared to an isolated heatmap because the justification of risk treatment does not remain in free text or implicit assumptions, but is present as model structure.

This paper focuses on methodological integration and structural correctness of the mapping steps. Nevertheless, certain aspects can be assessed systematically because they follow from model structure and process integration.
We consider five criteria that are practically relevant in both risk and audit contexts:
\begin{itemize}
  \setlength{\itemsep}{0.5em}
  \item \textbf{Transparency:} Are assumptions, context states, and thresholds documented explicitly?
  \par\smallskip\noindent The presented approach makes the modeling assumptions explicit by storing context factors/states and the prioritization thresholds as project artifacts that can be inspected and versioned. The Bowtie model (threats, barriers, consequences, and activation rules) is persisted as structured XML, and the generated DAG includes the resulting nodes, edges, and probability tables. This means an auditor can verify exactly which context and threshold settings led to a specific prioritization and network.
  
  \item \textbf{Traceability:} Is there a traceable chain from prioritization $\rightarrow$ causes/barriers $\rightarrow$ model structure?
  \par\smallskip\noindent Our approach maintains a traceable chain by linking heatmap-based prioritization to the selected Bowtie and then transforming that Bowtie into a DAG via an explicit, file-based pipeline in the implementation (Bowtie $\rightarrow$ intermediate model $\rightarrow$ DAG artifacts). Stable identifiers and names are carried through the transformation so each DAG element can be traced back to its originating cause/barrier in the Bowtie. The tooling therefore supports backtracking from a ranked risk to the concrete causal structure and control rationale.
  \item \textbf{Actionability:} Are concrete intervention points (controls) represented in the model?
  \par\smallskip\noindent Controls are first-class elements: Barriers in the Bowtie are modeled explicitly and carried into the DAG as variables whose states directly affect the probability of the top event and consequences. Our approach additionally supports activation state handling for barrier-related DAG nodes, enabling operational interpretation rather than purely descriptive modeling. Our tooling resorts to REST APIs for exposing the state of the identified intervention points for monitoring and real-time updates.
  
  \item \textbf{Reproducibility:} Do identical inputs (context, Bowtie) deterministically lead to the same DAG?
  \par\smallskip\noindent Given identical inputs, our approach’s mapping and export steps follow deterministic transformation rules, producing the same DAG structure and artifacts on repeated runs. Inputs are persisted before every transformation, and the generation process does not rely on random sampling. This supports repeatable audits and consistent comparisons across iterations.
  
  \item \textbf{Process scalability:} is the approach suitable for many risks (broad) and few deep dives?
  \par\smallskip\noindent Our approach supports broad coverage by using polar heatmaps to screen and prioritize many risks across multiple contexts with low modeling effort. For selected high-priority items, the same workflow supports deep dives by modeling a Bowtie and automatically deriving a DAG for monitoring and inference. The combination of reusable artifacts, automated transformations, and consistent tooling enables scaling across both portfolios and detailed investigations.
\end{itemize}

\subsection{Comparison to alternatives}

\textbf{Classical 2D risk matrix without context.} This alternative scales procedurally, but fails under context changes: the same risk description can lead to very different decisions in different operating states without the reason being visible in the artefact. The polar heatmap compensates for this by modelling context dimensions as axes and documenting them in each slice~\cite{hermann2026polar}. This addresses a central criticism of risk matrices, i.\,e., the lack of informative resolution towards \enquote{contextual resolution}~\cite{cox2008}.

\textbf{Bowtie without upstream context-based prioritization.} Bowties are explanatory but expensive. Without prioritization, organizations will either model too many Bowties (cost) or too few (blind spots). The proposed pipeline uses the polar heatmap as a filter and simultaneously provides the context parameters that are reused in the Bowtie structure. This reduces modeling effort and increases consistency between prioritization and deep dive.

\textbf{Quantitative models without structured barrier representation.} Alternatively, availability or incident models can be built directly as a BN or Markov model. In practice, however, an explicit, auditable representation of barriers and their place in the causal chain is often missing. Bowties and explicit barrier semantics provide an established framework~\cite{dianous2006}. Bowtie is also categorized as a risk analysis technique in ISO/IEC~31010~\cite{iec31010}. Through DAG transfer, this framework is transformed into a probabilistic model without losing the barrier view~\cite{khakzad2013}. Practical implementations of this transformation have already shown feasibility~\cite{zurheide2021pybnbowtie}.

\subsection{Advantages of the proposed approach}
In the DORA context, the combination of context sensitivity and control evidence is central. The proposed pipeline provides:
\begin{itemize}
  \item contextual prioritization based on polar heatmaps,
  \item control-oriented structure using Bowtie with barriers,
  \item evidence and intervention points at the DAG level.
\end{itemize}
Thus, the approach is not merely a visualization or modeling tool, but a bridge between risk governance, engineering controls, and (future) quantitative effectiveness analysis.

\section{Proposed Risk and Decision-Making Process}

The pipeline presented is not only a modeling approach, but can be integrated as a repeatable process step into existing governance structures. Especially in the DORA context, value is created when artifacts are not treated as \enquote{one-off documentation}, but as living models that absorb learnings from changes and incidents.

\subsection{Proposed workflow for integration}
A practicable workflow for teams (risk, SRE/operations, architecture) can look as follows:
\begin{enumerate}
  \item \textbf{Initial classification:} The risk is recorded as a heatmap item, baseline values $X_{\text{base}},Y_{\text{base}}$ are documented.
  \item \textbf{Context definition:} Relevant context axes are defined; ND slices for typical operating states are maintained (e.\,g., off-peak/peak, change window, third-party status).
  \item \textbf{Prioritization:} The polar heatmap serves as a review artifact; risks in the non-acceptable area are marked for deep dive.
  \item \textbf{Bowtie deep dive:} Causes, consequences, and barriers are modeled; barriers are classified (preventive/detective/mitigative) and marked as activation nodes.
  \item \textbf{DAG export:} The structure is transferred into a DAG; relevant telemetry sources are identified as potential evidence for the respective DAG nodes.
  \item \textbf{Review and action cut:} Results are discussed in risk board/change advisory; measures are prioritized based on barrier position and (later) sensitivity analyses.
\end{enumerate}

\subsection{Lifecycle: learnings from incidents and tests}
A central advantage of the model-based approach is reuse. After an incident or resilience test, insights can be captured not only as post-mortem text but as an update in the Bowtie/DAG (e.\,g., new cause, weak barrier, new dependency). Over time, this creates a knowledge artifact that accelerates and makes the next assessment more consistent.

\subsection{Pragmatic limits in operation}
For operation, it is important to control effort. The pipeline is therefore explicitly designed so that it can be used \emph{gradually}: many risks remain at heatmap level, only a few material risks are analyzed in depth. The decision when to deepen the analysis remains a governance rule (e.\,g., above a certain color class or for certain services). The polar heatmap provides a context-sensitive trigger for exactly this.

\section{Conclusion and outlook}
This paper describes an end-to-end pipeline from context-sensitive heatmap prioritization (polar heatmap) via Bowtie deepening to DAG-based operationalization as a basis for Bayesian networks. The implemented tools have been evaluated by modeling a DORA-motivated use case. The steps have been documented in this paper to show that context factors from the heatmap model are transferred consistently into Bowtie barriers and consequently DAG nodes that are marked as activation nodes. Future work includes (i) parameterization and estimation of probabilities as well as barrier effectiveness, (ii) quantitative intervention analyses (activation nodes as do-interventions), and (iii) continuous control monitoring.
For more information about the implemented tools and the integrated risk management platform, feel free to email us at \url{info@kompilomat.com}. 

\section*{AI Disclosure}
As our scientific work focuses on the integration of AI-based methods into various types of processes, particularly for security-related applications, large language models were used to support the revision of this manuscript. Their use was limited to editorial assistance, including the refinement of figures as well as improvements to clarity, wording, and error correction in the main text. The authors take full responsibility for the content of this work.

\bibliographystyle{IEEEtran}
\bibliography{references}

\end{document}